\documentclass[aps,showpacs,pra,reprint]{revtex4-1}

\usepackage[english]{babel} 
\usepackage[ansinew]{inputenc} 
\usepackage[T1]{fontenc}	
\usepackage[]{graphicx} 
\usepackage{amsmath} 
\usepackage{dsfont} 
\usepackage[colorlinks=true,allcolors=blue]{hyperref}

\usepackage[normalem]{ulem} 
\usepackage[usenames,dvipsnames]{color} 
\definecolor{orange}{rgb}{0.7,0.2,0}
\definecolor{darkgreen}{rgb}{0,0.3,0}

\newcommand{\bra}[1]{\ensuremath{\left\langle #1\right|}}
\newcommand{\ket}[1]{\ensuremath{\left|#1\right\rangle}}

\begin{document}
\title{Derivation of Markovian master equations for spatially correlated decoherence}  
\author{Jan Jeske and Jared H. Cole}
\affiliation{Chemical and Quantum Physics, School of Applied Sciences, RMIT University, Melbourne, 3001, Australia}

\begin{abstract}
We introduce a general formalism to describe the effects of Markovian noise which is spatially correlated, typically decaying over some finite correlation length. For any system of interest, this formalism describes spatial correlations without the necessity to choose a particular microscopic model for the environment. We present a method of mapping the equations to Lindblad form and discuss functional forms for homogeneous spatial correlation functions. We also discuss two example microscopic models for the environment which exhibit non-trivial spatial-temporal correlation functions.
\end{abstract}

\maketitle
\section{Introduction}
The study of open quantum systems and the concept of a density matrix master equation underlies much of modern quantum physics, be it quantum optics, atom optics, condensed matter physics or quantum computation.  When deriving or assuming a particular form for the master equation via the usual system-bath model \cite{Breuerbook}, it is common to assume each component of a system either couples to the same bath (\emph{correlated} or \emph{collective} decoherence channels) or individual baths (\emph{uncorrelated} or \emph{independent} decoherence channels), see e.g.~\cite{Terhal2005, Doll2007, Contreras2008, McCutcheon2009, Jeske2012}. We explore the regime between these two extremes, introducing the concept of a \emph{correlation length} $\xi$ and deriving a general master equation method for treating such partially correlated environments.

Decoherence induced by a correlated environment is a commonly observed effect.  The concepts of super- and sub-radiance \cite{Dicke1954, Crubellier1985} and decoherence-free subspaces \cite{Lidar1998, Lidar2001, Lidar2001_II, Lidar2003, Blume-Kohout2008, Doll2006} rely on the interference resulting from several subsystems coupled to the same environmental mode.  Even some of the foundational work on decoherence in quantum computation considered both fully correlated and uncorrelated environments \cite{Palma1996, Reina2002, Duan1998, Zanardi1998}.  Recently, the ramifications of correlated environments have been discussed in such diverse situations as scalable quantum error correction \cite{Alicki2002, Averin2003, Clemens2004, Aharonov2006, Terhal2005, Aliferis2006, Novais2006}, photosynthesis and biological chromophores \cite{Caruso2009, Scholes2012, Yang2002, Palmieri2009, Ringsmuth2012} and multi-atom trapping experiments \cite{Safavi-Naini2011, Monz2011, Deslauriers2006, Schroll2003}.

The Lindblad equation \cite{Lindblad1976, Gorini1976, Carmichaelbook, Breuerbook} is the workhorse of open quantum systems due to its simple form and well behaved mathematical properties.  Yet, deriving a general master equation of this form for (partially) correlated environments is non-trivial.  Lindblad operators can be derived or assumed which act individually, pair-wise or collectively, yet how does one choose these in a physically sensible manner? We use a general Bloch-Redfield approach, where environmental noise correlation functions appear naturally in the formalism. Given sufficiently well behaved environmental correlations, a closed form master equation can be obtained with the same form as the Lindblad equation but whose operators and rates are linked directly to the original physical system-bath Hamiltonian.  Following this route we consider a generalisation of the environmental correlation function which includes spatial (as well as temporal) correlations and therefore derive a general master equation. This describes spatially correlated decoherence independent of the particular bath Hamiltonian and purely based on environmental correlation functions. We also consider several examples where such \emph{spatial-temporal} correlation functions can be derived microscopically.

\section{Bloch-Redfield equations with spatial correlations}
Starting from the usual system-bath Hamiltonian,
\begin{align}
H=H_{S}+H_{B}+H_{int} \label{Hamiltonian parts}
\end{align}
comprised of system ($H_S$), bath ($H_B$) and the interaction between them
\begin{align}
H_{int}=\sum_j s_j B_j \label{interaction Hamiltonian}
\end{align}
where $s_j$ are system operators and $B_j$ bath operators. 
An arbitrary basis $\{ \ket{a_n} \}$ and the Hamiltonian eigenstates $H_S \ket{\omega_n}= \omega_n \ket{\omega_n}$ are connected via the transformation matrix $V=\sum_n \ket{\omega_n} \bra{a_n}$. A compact and general form of the Bloch-Redfield equations is then given by (see appendix \ref{app B-R derivation}):
\begin{multline}
\dot \rho = \frac{i}{\hbar} [\rho,H_s] + \frac{1}{\hbar^2} \sum_{j,k} \left(-s_j V q_{jk} V^\dagger \rho + V q_{jk} V^\dagger \rho s_j 
\right.\\ \left. 
-\rho V \hat q_{jk} V^\dagger s_j + s_j \rho V \hat q_{jk} V^\dagger\right) \label{B-R equations}
\end{multline}
\begin{align}
&\text{with} \nonumber\\
&\langle a_n| q_{jk} |a_m\rangle = \bra{a_n} V^\dagger s_k V \ket{a_m} \frac{1}{2} C_{jk}(\omega_m-\omega_n) \label{q-elements}\\
&\langle a_n| \hat q_{jk} |a_m\rangle = \bra{a_n} V^\dagger s_k V \ket{a_m} \frac{1}{2} C_{kj}(\omega_n - \omega_m) \label{qhat-elements}\\
&C_{jk}(\omega)= \int_{-\infty}^\infty d \tau \; e^{i \omega \tau} \, \langle  \tilde B_j(\tau) \tilde B_k(0) \rangle \label{spectral function definition}
\end{align}
The spectral functions $C_{jk}(\omega)$ define the bath, i.e.~the environment. They are given by the Fourier transform of the correlation functions of the bath operators in the interaction picture $\tilde B_j(\tau)=\exp(i H_B \tau) B_j \exp(-i H_B \tau)$. 

\subsection{Spatial-temporal correlations}
The spectral functions define both the correlation function with increasing time difference $\tau$ and which of the pairs of bath operators $B_j, B_k$ (corresponding to system operators $s_j, s_k$) are correlated. The operators $B_j$ can be grouped such that 
\begin{align*}
C_{jk}(\omega)=\left\{ 
\begin{aligned} &C_{jk}(\omega) \;\; &&\text{ if $j$ and $k$ are in the same group}\\
&0\;\; &&\text{ if $j$ and $k$ are in different groups}  \end{aligned} \right.
\end{align*}
This may be due to either the assumption of a certain structure in the environment or groups of different types of coupling operators (see appendix \ref{app secular approximation of B-R}). We call each group an independent bath since they are not correlated with the other baths. 

If the environment is thought to be a continuum with decaying correlations over increasing distances a more flexible model is to introduce a spatial dependency in the spectral function:
\begin{align}
C_{jk}(\omega, \textbf{r}_j, \textbf{r}_k):= \int_{-\infty}^\infty d \tau \; e^{i \omega \tau} \, \langle  \tilde B_j(\tau, \textbf{r}_j) \tilde B_k(0, \textbf{r}_k) \rangle
\end{align}
where the components of the system couple to the environment at positions $\textbf{r}_j, \textbf{r}_k$. This function is similar to a Van-Hove-function \cite{vanHove1954}, however we use real-space coordinates rather than k-space.
As in conventional Bloch-Redfield theory, this spectral function can be derived from a more fundamental microscopic model or its form can be phenomenologically assumed, typically as a homogeneous spectral function $C_{jk}(\omega,\textbf{r}_j, \textbf{r}_k)=C(\omega, |\textbf{r}_j - \textbf{r}_k|)$ which will approach zero for increasing distance.

Strictly speaking the spectral function $C_{jk}(\omega)$ is actually given by a one-sided Fourier transform $D_{jk}(\omega)=C_{jk}(\omega)+i F_{jk}(\omega)$, i.e.~it may be complex. Such a term would lead to additional coherent dynamics. In the secular approximation it can be written as a correction $H_S \rightarrow H_S + H_{cor}$ to the system Hamiltonian (see eq.~\ref{B-R with correction Hamiltonian}), where $H_{cor}=\sum_{j k } F_{jk} s_k^\dagger s_j$. In cases of uncorrelated decoherence (all terms for which $j\neq k$ are zero) this correction of the system Hamiltonian is generally neglected since the terms $s_j^\dagger s_j$ in qubit systems are diagonal in the system eigenbasis. In cases of correlated decoherence however the correction Hamiltonian $H_{cor}$ can lead to interaction terms between the qubits, i.e.~environmentally induced interactions.

\subsection{Qubit example}
\label{sec qubit example}
As an example we apply this formalism to two uncoupled qubits $H_S=\sum_{j=1}^2\omega_q\sigma_z^{(j)}$, each interacting longitudinally with the environment $H_{int}=\sum_{j=1}^2\sigma_z^{(j)}B_j$. With the four states \ket{1,1},\ket{1,0},\ket{0,1},\ket{0,0} we find a reduced dephasing rate $\gamma_-$ for the single excitation subspace $\{\ket{1,0},\ket{0,1}\}$ and an enhanced dephasing rate $\gamma_+$ between the states \ket{1,1} and \ket{0,0} while all other pairs dephase at a rate $\gamma_0$. These rates are obtained in terms of $C(\omega, |\textbf{r}_j - \textbf{r}_k|)$ as,
\begin{align}
\gamma_- &= C(0, 0)-C(0, d) \label{gamma_red}\\
\gamma_+ &= C(0, 0)+C(0, d)\label{gamma_enh}\\
\gamma_0 &= C(0, 0)/2
\end{align}
where $d=|\textbf{r}_1 - \textbf{r}_2|$ is the distance between the qubits. For uncorrelated decoherence only the self-correlations are non-zero and all coherences decay at the rate $\gamma_0\;\text{or}\;2\gamma_0$. With increasing noise correlation length and fixed qubit distance the single excitation subspace's dephasing rate $\gamma_-$ is reduced while $\gamma_+$ is increased. This reduction of $\gamma_-$ is the basis of a decoherence-free subspace\cite{Lidar1998}. 

For $n$ qubits in an uncorrelated environment the dephasing rate between two states is proportional to the number $n_f$ of flipped qubits between the two states. In a perfectly correlated environment however the dephasing rate between two states with a difference of $n_e$ excitations is proportional to $n_e^2$ and $n_f$ is irrelevant. Therefore the dephasing rate between states with equal excitation number is reduced to zero when the noise correlation length increases well beyond the qubits' separation, forming a decoherence-free subspace. For example a coherence of the form \ket{0011}\bra{1100} will decay with rate $\Gamma = n_f \gamma = 4\gamma$ for $\xi \rightarrow 0$ and as $\Gamma = n_e^2 \gamma = 0$ for $\xi \rightarrow \infty$, where $\gamma$ is the corresponding single qubit dephasing rate. In contrast, the coherence \ket{0000}\bra{1111} which also decays as $\Gamma = n_f \gamma = 4\gamma$ for $\xi \rightarrow 0$, will decay as $\Gamma = n_e^2 \gamma = 16 \gamma$ for $\xi \rightarrow \infty$, i.e.~the rate increases immensely for long correlation length.

%

\section{Analytical spatial correlation functions \label{sec analytical spatial correlation function}}
To illustrate the concept of spatially correlated decoherence, we now consider two example microscopic models for an environment without choosing a particular system Hamiltonian. The environmental spatial-temporal correlation function as well as the spectral function can be calculated explicitly showing different spatial correlations in each example. In the first case the correlation length links naturally to the environmental parameters, in the second case the spatial correlations oscillate with increasing distance.

\subsection{One-dimensional Ising model}
\label{sec ising model}
The first microscopic example is one of a system dephasing due to the influence of a classical one dimensional Ising chain of $N$ coupled spins $S_x=\pm 1$ with Hamiltonian:
\begin{align}
H_B=-J \sum_{x=1}^{N-1} S_x S_{x+1}
\end{align} 
with coupling strength $J$. The spatial correlations in a thermal equilibrium state are given by an exponential decay \cite{Noltingmagnetism}:
\begin{align}
\langle S_x S_{x'} \rangle = \tanh^{|x-x'|}(\beta J) = \exp\left\{\ln [\tanh( \beta J)] \; |x-x'| \right\}\label{correlation function (space) for 1D Ising model}
\end{align}
with $\beta = 1/k_B T$. This can be extended \cite{Glauber1963, Kawasaki1972} to a spatial-temporal correlation function by introducing a switching rate per unit time $\alpha/2$ between the two states $S_x=\pm 1$ for each spin (due to a heat bath), leading to an infinite sum of exponential decays in space multiplied with modified Bessel function of the first kind $I_n(\tau)$ in time:
\begin{align}
\langle S_x(0)S_{x'}(\tau)\rangle= \sum_{l=-\infty}^\infty \eta^{|x-x'+l|} I_l(\gamma \alpha |\tau|)\; e^{-\alpha |\tau|} \label{correlation function (space-time) for 1D Ising model}
\end{align}
where $ \eta=\tanh(\beta J);$ and $\gamma=\tanh(\beta 2J) $. Note that setting $\tau \rightarrow 0$ in eq.~\eqref{correlation function (space-time) for 1D Ising model} yields eq.~\eqref{correlation function (space) for 1D Ising model}.


Dephasing in qubit systems is generally caused by the noise at zero-frequency and the spatial correlations at $\omega=0$ can be calculated (assuming positive $J$) by integrating over each summand in eq.~\ref{correlation function (space-time) for 1D Ising model} and evaluating the sum afterwards,
\begin{align}
C(\omega=0,|x-x'|) = \frac{2 \left(|x-x'| +\zeta \right) \zeta \eta^{|x-x'|} }{\alpha },
\end{align}
with $\zeta=\cosh(2 J \beta )$ and $ \eta=\tanh(\beta J)$. The spectral function decays with distance as $x \exp(-x)$ and with it the collective terms of the dephasing rate (cf.~eq.~\eqref{gamma_red} and \eqref{gamma_enh}) decay.  The effective correlation length in this example increases with the coupling $J$ of the environmental spins and decreases with temperature $k_B T$. Note that the spatial indices here are natural numbers, i.e.~in units of the separation between environmental Ising spins, $w$. 

Using this spectral function for our previous example of a two qubit system (section \ref{sec qubit example}) one finds the rates
\begin{align}
\gamma_- &= \frac{2 \zeta}{\alpha } \left[ \zeta - \eta^d(d+\zeta) \right]\\
\gamma_+ &= \frac{2 \zeta}{\alpha } \left[ \zeta + \eta^d(d+\zeta) \right]\\
\gamma_0 &= \zeta^2/\alpha
\end{align}
where $\zeta=\cosh(2 J \beta )$ and $ \eta=\tanh(J \beta);$ and the qubit distance $d$ is given in units of the environmental Ising spins' nearest-neighbour distance. For close distances $d\ll J \beta$ the correlated decoherence effects are the strongest and the reduced dephasing rate $\gamma_-$ is close to zero.  

\subsection{Bosonic chain}
One of the most studied system-bath models is the spin-boson model \cite{Leggett1987, DiVincenzo1995, Unruh1995, Palma1996, Zanardi1998} where a spin undergoes dephasing due to an ensemble of uncoupled harmonic oscillators.  To generalise this model, we now ask the question of how two (or more) spins, located at positions $x$ and $x'$ (etc.), are affected by a common bath with non-trivial correlations. As an example of such a bath we consider a one-dimensional chain of spatially located, coupled harmonic oscillators, the `tight-binding chain', described by:
\begin{align}
H_B=\sum_{x=1}^N \omega_0 a_x^\dagger a_x - g \sum_{x=1}^{N-1} \left( a_x^\dagger a_{x+1} + a_{x+1}^\dagger a_x \right) \label{Hamiltonian of coupled bosonic chain in space}
\end{align}
Transforming into $k$-space via a lattice Fourier transform \cite{Noltingmanybody} diagonalises the Hamiltonian for large $N$, $H_B=\omega_k \sum_k a_k^\dagger a_k$ with $\omega_k = \omega_0 -2 g \cos k w$ where $w$ is the environmental lattice spacing.  For a finite chain length, the sum over k is given by $k_n = 2\pi n/Nw$ for integer $n\in[-N/2,N/2]$.

We choose the bath operators $B_j$ in the interaction Hamiltonian (eq. \eqref{interaction Hamiltonian}) to be the lowering and raising operators $a_x, a_x^\dagger$ at a particular position $x$, 
\begin{align*}
B(x)=a_x = \sum_k \frac{e^{i k x w}}{\sqrt{N}} a_k , \quad  B^\dagger(x)=a_x^\dagger  = \sum_k \frac{e^{-i k x w}}{\sqrt{N}} a_k^\dagger 
\end{align*}
which are non-local in $k$-space \cite{Noltingmanybody} and similar in form to ref.~\cite{Zanardi1998}.  
\subsubsection{Spatial-temporal correlation functions}
If we assume the steady-state density matrix of the harmonic oscillator chain is a fully mixed state (in the Fock basis of the k-modes) we can compute the four spatial-temporal correlation functions for a coupled chain of harmonic oscillators:
\begin{align}
\langle\tilde B(\tau,x) \tilde B(0,x')\rangle = \langle \tilde B^\dagger(\tau,x) \tilde B^\dagger(0,x')\rangle= 0
\end{align}
and
\begin{align}
\langle \tilde B^\dagger(\tau,x) \tilde B(0,x') \rangle &= \sum_k  \frac{1}{N} e^{-i k (x-x') w} e^{i \omega_k \tau} \langle n_k \rangle \label{correlation function of bosonic environment}\\
\langle \tilde B(\tau,x) \tilde B^\dagger(0,x') \rangle &= \sum_k \frac{1}{N} e^{i k (x-x') w} e^{-i \omega_k \tau} (1+\langle n_k \rangle)\label{correlation function of bosonic environment 2}
\end{align}
where $\langle n_k\rangle$ is the expectation value of the occupation number of mode $k$.   These expressions are exact and can be evaluated numerically for finite $N$.  To obtain more physical insight we take the limits of large energies, where $\langle n_k\rangle \approx \exp(-\hbar \omega_k \beta)$, and a long chain
\begin{align}
\sum_{n=-N/2}^{N/2} f(k_n) \rightarrow \frac{Nw}{2\pi} \int_{-\pi /w}^{\pi/w} dk f(k).
\end{align}
If we assume that the correlations in the chain are dominated by propagating excitations, we may also linearise the dispersion relation, $\omega_k \approx \omega_0 + 2g(|k|w-\pi/2)$ within the first Brillouin zone.

Taking these limits, we find that the correlation function decays for large $\tau$ as
\begin{align}
\lim_{\tau \rightarrow \infty} \langle \tilde B^\dagger(\tau,x) \tilde B(0,x') \rangle  \propto \frac{1}{2 \pi g \tau}.
\end{align}
This slow algebraic temporal decay indicates the potential for non-Markovian memory effects of this environment. However, the coupling strength $g$ is an environmental parameter and in the limit where $g$ is much larger than the relevant system parameters the decay time of correlations is much shorter than the relevant system time scales. Only in this limit the corresponding spectral function can be used for Markovian master equations.

For linearised dispersion, we can also evaluate the spatial correlations at one moment in time $t=t' \Leftrightarrow \tau=0$, in the limit of $g \beta \gg 0$ and normalised by the value for $x=x'$,
\begin{align}
\frac{\langle \tilde B^\dagger(t,x) \tilde B(t,x') \rangle}{\langle \tilde B^\dagger(t,x) \tilde B(t,x) \rangle} &=\frac{(2\beta g)^2}{(2\beta g)^2+(x-x')^2}.\label{relative spatial corfunc at tau=0 for bosonic chain}
\end{align}
This indicates that the mixed spatial correlations decay over a characteristic correlation length $\xi = 2 w \beta g$, due to the interplay between excitations hopping along the chain and thermal noise.

\subsubsection{Spectral functions}
While the relative spatial correlation function for $\tau=0$ shows Lorentzian decay (eq.~\ref{relative spatial corfunc at tau=0 for bosonic chain}) the spectral function shows very different spatial correlations. The spectral function of eq. \eqref{correlation function of bosonic environment} can be computed in the limit of large $N$, using the identity $\delta \left( f(k) \right) = \sum_{j} \frac{\delta(k-k_j)}{|f'(k_j)|}$ where $f(k_j)=0$. For cosine dispersion and a Boltzmann distribution for $\langle n(\omega)\rangle$ we find
\begin{multline}
C_{B^\dagger B}(\omega,x-x')= \Theta\left(2 |g| - |\omega+\omega_0 |\right)\\
\cdot \frac{
\cos\left[ (x-x') \arccos \left(-\frac{\omega+\omega_0}{2 g} \right)  \right]
\langle n(|\omega|) \rangle}{\pi \sqrt{4 g^2 - (\omega + \omega_0)^2}}  
\label{spectral function for bosonic coherent chain (cosine, Boltzmann distr)}
\end{multline}
The spectral function $C_{BB^\dagger}(\omega,x-x')$ corresponding to eq.~\eqref{correlation function of bosonic environment 2} takes the same form with $\omega \rightarrow -\omega$ and $\langle n(|\omega|)\rangle \rightarrow (1+\langle n(|\omega|)\rangle)$.

The spectral functions show that the noise consists of frequencies centered around the oscillator's eigenenergies $\omega_0$ with a noise cut-off at $\omega_0 \pm 2 g$ due to the band gap of the chain. For a given frequency the spatial correlations are cosine-oscillations with distance $|x-x'|$, allowing points of negative correlations. The cosine shape is physically necessary as the self-correlations (at $|x-x|=0$) always need to be positive. 

With this spectral function the rates in our previous example of a two qubit system (section \ref{sec qubit example}) become
\begin{align}
\gamma_- &= [1-\cos (d \pi/2)]/(2 \pi g \langle n(0)\rangle)\\
\gamma_+ &= [1+\cos (d \pi/2)]/(2 \pi g \langle n(0)\rangle)\\
\gamma_0 &= 1/(\pi g \langle n(0)\rangle)
\end{align}
where $d$ is given in units of the environmental nearest-neighbour distance of the harmonic oscillators $w$ and we assume $\omega_0 \approx 0$ on the scale of the system energies making their energy spectrum approximately continuous. There are now three types of system dynamics: For $\cos(d \pi/2)=1$ one finds the qubits fully (positively) correlated and $\gamma_- \rightarrow 0$, similar to the previous model (section \ref{sec ising model}) for short distances. At the points where $\cos(d \pi/2)=0$ the collective terms are zero and all correlated decoherence effects vanish. For $\cos(d \pi/2)=-1$ the collective terms are negative and $\gamma_+ \rightarrow 0$, i.e.~the reduced and enhanced rates swap roles. This means the subspace $\{\ket{1,1}, \ket{0,0} \}$ becomes decoherence-free and the subspace $\{\ket{1,0},\ket{0,1}\}$ has an enhanced dephasing rate. This rare case only occurs at points where the environmental noise of the two qubits is negatively correlated.

\section{Phenomenological spatial correlation functions and valid functional forms}
Ideally one has a clear microscopic model of a particular environment and can derive a spatial-temporal correlation function from it. However in some cases it is necessary to assume the spatial-temporal correlations phenomenologically. We now present a method to map the Bloch-Redfield equations to Lindblad form, which provides a test of complete positivity in the time evolution for any such phenomenological spatial-temporal correlation function. We furthermore present two examples of valid functional forms.
\subsection{Mapping to Lindblad equations}
The Bloch-Redfield equations do not guarantee complete positivity by their mathematical form but depend on a consistent model of the environment's temporal and spatial correlation functions. When assuming a phenomenological spatial-temporal correlation function $C_{jk}(\omega, \textbf{r}_j, \textbf{r}_k)$, highly nontrivial physical conditions apply for multipartite correlations. We therefore present a way to check for a mapping to the Lindblad equations which are known to be ``the most general form of the generator of a quantum dynamical semigroup'' \cite{Breuerbook, Gorini1976, Lindblad1976}. This means that the time-evolution operator due to a certain Markovian master equation is a completely positive map of the density matrix if and only if there exists a set of operators in which the master equation takes on Lindblad form. 

We take the Bloch-Redfield equations in the secular approximation (see appendix \ref{app secular approximation of B-R}):
\begin{align}
 \dot \rho = \frac{i}{\hbar} [\rho,H_s] + \frac{1}{\hbar^2} \sum_{j,k} C_{jk}  \left(  s_k \rho s_j^\dagger - \frac{1}{2} \left\{ s_j^\dagger s_k, \rho \right\} \right) \label{Lindblad non-diagonal form}
\end{align}
This equation can be mapped to Lindblad form if and only if there exists an invertible matrix $W$ such that $W^\dagger C_{jk} W$ is diagonal with non-negative eigenvalues. The eigenvalues then correspond to the Lindblad rates. The Lindblad operators are given by:
\begin{align}
L_k=\sum_j W_{jk} s_j 
\end{align}
For real-valued homogeneous correlation functions, $C_{jk}(\textbf{r}_j, \textbf{r}_k)=C_{jk}(|\textbf{r}_j - \textbf{r}_k|)$, the coefficient matrix $C_{jk}$ is symmetric and its diagonalisation is unitary, i.e.~eq.~\eqref{Lindblad non-diagonal form} can be mapped to Lindblad form if the coefficient matrix is positive semi-definite. If one or more eigenvalues are negative then according to Sylvester's law of inertia \cite{SchejaStorchbook} all transforms $W^\dagger C_{jk} W$ will have the same number of negative values on the diagonal, i.e.~there is no mapping to Lindblad form.

\subsection{Valid functional forms}
To illustrate this procedure, we consider a one-dimensional chain of equidistant subsystems (separated in space by a distance $d$). We now show that the following two functional forms can always be mapped to Lindblad form. Both are therefore sensible physical models of spatial correlations which guarantee complete positivity in the time evolution:
\begin{align}
C_{jk}&=\exp[-a|j-k|] &&\text{exponential decay} \label{exponential decay correlation function}\\
C_{jk}&=\exp[-a (j-k)^2] &&\text{Gaussian function} \label{Gaussian correlation function}
\end{align}
where the parameter $a>0$ sets the decay in terms of the environmental correlation length, $a = d/\xi$.  

We first note that homogeneous coefficient matrices $C_{jk}$ with row $j$ and column $k$ are Toeplitz-matrices. To determine whether it is positive semi-definite we utilise Lemma 6 in Chapter 4.2 of reference \cite{Graybook}, which applied to our notation reads:

The eigenvalues of an $n$x$n$ Hermitian Toeplitz matrix in the notation
\begin{align}
(C_{jk})=
\left( \begin{array}{cccccc} 
 t_0 & t_{-1} & t_{-2} & t_{-3} & \dots & t_{-n+1}\\
 t_1 & t_{0}  & t_{-1} & t_{-2} & \ddots &\\
 t_2 & t_1    & t_0    & t_{-1} & \ddots &\\
 t_3 & t_2    & t_1    & t_0    & \ddots &\\
 \vdots &\ddots & \ddots & \ddots &\ddots & \vdots\\
 t_{n-1} & & & &\dots &t_0
\end{array} \right) \label{Toeplitz matrix notation}
\end{align}
with absolutely summable elements $t_m$,
\begin{align}
\sum_{m=-\infty}^\infty |t_m| < \infty
\end{align}
are not smaller than the minimum and not greater than the maximum of the Fourier series $f(\lambda)$ defined by
\begin{align}
f(\lambda)= \sum_{m=-\infty}^\infty t_m e^{-i m \lambda}.
\end{align}

\subsubsection{Exponential decay}
Since the element index in eq. \ref{Toeplitz matrix notation} corresponds to the difference (row - column) the Toeplitz matrix corresponding to eq. \ref{exponential decay correlation function} is given by the elements $u_m=e^{-a|m|}$.  Via a geometric series, they are found absolutely summable:
\begin{align}
\sum_{m=-\infty}^{\infty} |u_m| = 1 +2\sum_{m=1}^\infty e^{-a m} = \frac{2}{1-e^{-a}}-1 < \infty
\end{align}
Similarly the Fourier series evaluates to:
\begin{align}
g(\lambda) & = \sum_{m=-\infty}^\infty u_m \exp(-i m \lambda) \\
		  & = -1+\frac{2 (1-e^{-a} \cos \lambda)}{1-2e^{-a} \cos \lambda +e^{-2a}} > 0
\end{align}
Since the Fourier series $g(\lambda)$ is greater than zero, the eigenvalues of the corresponding Toeplitz matrices are greater than zero. Any master equation based on an exponentially decaying spatial correlation function $C_{jk}=\exp(-a|j-k|)$ can therefore be mapped to Lindblad form.

\subsubsection{Gaussian decay}
The Toeplitz matrix corresponding to eq. \ref{Gaussian correlation function} has absolutely summable elements $v_m=e^{-a m^2}$:
\begin{align}
\sum_{m=-\infty}^\infty |v_m| = 2 \sum_{m=0}^\infty e^{-a m^2} -1 <\infty 
\end{align}
while the Fourier series is given by,
\begin{align}
h(\lambda)&=\sum_{k=-\infty}^\infty v_k e^{ik\lambda} = 1+2\sum_{k=1}^\infty e^{-ak^2} \cos(k\lambda) > 0
\end{align}
which can be shown using Jacoby's imaginary transformation (chapter 21, page 475 of reference \cite{Whittaker1963}).
Again we see that the Fourier series $h(\lambda)$, and therefore the eigenvalues of the corresponding Toeplitz matrices are greater than zero which means that any master equation based on a Gaussian correlation function can be mapped to Lindblad form.

\subsubsection{Step function}
Finally we give a counterexample where the mapping fails, indicating an unphysical correlation function. The Toeplitz matrix corresponding to a step correlation function has the elements:
\begin{align}
w_m=\left\{ \begin{aligned} 1  &\quad \text{for }|m|<2 \\ 0 &\quad \text{otherwise} \end{aligned} \right.
\end{align}
The elements are absolutely summable $\sum_{m=-\infty}^{\infty} |w_m|=3$ however the Fourier series has negative values:
\begin{align}
\sum_{k=-\infty}^\infty w_k e^{ik\lambda} = e^{i \lambda} + e^{-i\lambda} +1 = 1+2\cos \lambda
\end{align}
In fact for the specific case of a system consisting of three subsystems
\begin{align}
\det \left( \begin{array}{ccc} 
1&1&0\\ 
1&1&1\\
0&1&1 
\end{array} \right) =-1, \label{step function matrix}
\end{align}
i.e.\ the matrix is not positive semi-definite. This 3x3 matrix is a principal submatrix \cite{Pemberton2001} of the coefficient matrix for any correlation function which corresponds to a step function. Following the definiteness criterion of symmetric matrices \cite{Pemberton2001, Jonesbook}, it follows that no step function correlation matrix (other than trivial limiting cases) is positive semi-definite. Step functions should therefore not be used as the form of a \emph{homogeneous} spatial correlation function as they always lead to a time evolution which is not completely positive, 
even though the temporal component can be considered to be Markovian.

\section{Conclusion}
We have utilized the Bloch-Redfield formalism to include spatial correlations within a decohering environment.  This approach is very general as the effect of the environment is modelled via spatial-temporal correlation functions. 

The typical effects of spatially correlated decoherence on qubit systems  are the formation of decoherence-free subspaces as well as enhanced dephasing rates between other states. For positively correlated noise the dephasing rate between two states is proportional to the square of the excitation difference of the two states.

Two microscopic examples for an environment were presented where the spatial-temporal correlation function can be derived explicitly.  The one-dimensional Ising model has a purely positive spectral function which decays spatially. The spatial correlation length is increased by stronger coupling  between the environmental spins and decreased by thermal effects. The chain of harmonic oscillators shows a spectral function which oscillates spatially in the form of a cosine, creating points of negative correlation. 

We also provide a method for working with phenomenologically assumed spectral functions. A mapping to the completely positive Lindblad form master equation provides a test for physicality of the spectral functions in the Markovian regime. With this test we show that homogeneous exponential or Gaussian spatial decay with an arbitrary correlation length prove to be physical models for spectral functions.

\acknowledgments
We would like to thank the following people for valuable discussions: M.~Marthaler, C.~M\"uller and N.~Vogt about the Bloch-Redfield formalism, S.~Huelga about correlated decoherence and P.~Longo about bosonic systems.

\appendix

\section{Derivation of the matrix form of the Bloch-Redfield equations}
\label{app B-R derivation}
The derivation of the Bloch-Redfield equations \cite{Bloch1957, Redfield1957, Wangsness1953} is a well known procedure which can be found in several good textbooks \cite{Carmichaelbook, Breuerbook, Rivasbook}.  In this and the following appendices, we summarise this derivation and its associated results for two reasons, firstly to clarify our notation and form of the equations and secondly to discuss several important approximations explicitly.  Although these approximations are included in the usual discussions, the introduction of spatial correlations requires careful treatment of these points.

From the usual system-plus-bath approach:
$$H=H_{S}+H_{B}+H_{int} \quad \text{with} \quad H_{int}=\sum_j s_j B_j$$
where the interaction is given as a product of system operators $s_j$  and bath operators $B_j$ one finds the standard master equation in the Born-Markov-approximation (see e.g.~Carmichael \cite{Carmichaelbook} eq. 1.34):
\begin{align*}
\dot{\tilde \rho}=\sum_{jk} \int_0^\infty dt' \left[ \tilde s_j(t) \tilde s_k(t') \tilde \rho - \tilde s_k(t') \tilde \rho \tilde s_j(t) \right] \langle \tilde B_j(t) \tilde B_k(t')\rangle \\
+\sum_{jk} \int_0^\infty dt' 
\left[ \tilde \rho \tilde s_k(t') \tilde s_j(t) - \tilde s_j(t) \tilde \rho \tilde s_k(t') \right] \langle \tilde B_k(t') \tilde B_j(t) \rangle
\end{align*}
where the tilde implies the interaction picture. To define our matrix notation in an arbitrary basis $\{ \ket{a_n} \}$ we introduce the transformation matrix $V=\sum_n \ket{\omega_n} \bra{a_n}$ to the system Hamiltonian's eigenstates $H_S \ket{\omega_n}= \omega_n \ket{\omega_n}$. To perform the $t'$ integration explicitly, we transform back to the Schr\"odinger picture.  For the first term it reads explicitly
\begin{align}
s_j V \underbrace{V^\dagger e^{-i H_S \tau} s_k e^{i H_S \tau} V}_{=:Q} V^\dagger \rho(t)
\end{align}
with $\tau=t-t'$. Inserting unity operators $\sum_n \ket{\omega_n}\bra{\omega_n}$ gives:
\begin{align}
Q&=\sum_{lmnp}\ket{a_l} \underbrace{\bra{\omega_l}e^{-iH_S\tau}\ket{\omega_m}}_{=e^{-i \omega_l \tau}\delta_{lm}} \bra{\omega_m}s_k\ket{\omega_n} \underbrace{\bra{\omega_n}e^{i H_S \tau}\ket{\omega_p}}_{=e^{i\omega_n \tau}\delta_{np}} \bra{a_p}\\
&=\sum_{mn}\ket{a_m} \bra{\omega_m}s_k\ket{\omega_n}\bra{a_n} e^{i(\omega_n-\omega_m)\tau} 
\end{align}
\begin{align}
\bra{a_m}Q\ket{a_n}&=\bra{\omega_m}s_k\ket{\omega_n} e^{i(\omega_n-\omega_m)\tau} \\
&=\bra{a_m}V^\dagger s_k V \ket{a_n} e^{i(\omega_n-\omega_m)\tau}
\end{align}
Performing these operations analogously on all four terms we can then integrate element-wise and we find a compact and general form of the Bloch-Redfield equations:
\begin{align}
\dot \rho = \frac{i}{\hbar} [\rho,H_s] + & \frac{1}{\hbar^2} \sum_{j,k} \left(-s_j V q_{jk} V^\dagger \rho + V q_{jk} V^\dagger \rho s_j \right. \nonumber\\ 
&\left. -\rho V \hat q_{jk} V^\dagger s_j + s_j \rho V \hat q_{jk} V^\dagger \right) \label{B-R equations in matrix form appendix}
\end{align}
\begin{align}
&\text{with} \nonumber\\ 
&\langle a_n| q_{jk} |a_m\rangle = \bra{a_n} V^\dagger s_k V \ket{a_m} D_{jk}(\omega_m-\omega_n)\\
&\langle a_n| \hat q_{jk} |a_m\rangle = \bra{a_n} V^\dagger s_k V \ket{a_m} D_{jk}^*(\omega_n - \omega_m)\\
&D_{jk}(\omega)=\int_{0}^\infty d \tau \; e^{i \omega \tau} \, \langle  \tilde B_j(\tau) \tilde B_k(0) \rangle \label{Laplace transform definition}
\end{align}
If $D_{jk}(\omega)$ contains an imaginary term, this leads to additional coherent dynamics (which in the secular approximation can even be written as a correction of the system Hamiltonian (see section ``Complex spectral functions''),
therefore only the real part of $D_{jk}$ causes decoherence. Assuming the property $\langle  \tilde B_j(-\tau) \tilde B_k(0) \rangle = \left( \langle  \tilde B_j(\tau) \tilde B_k(0) \rangle \right)^*$ this real part can be rewritten in terms of the Fourier transform:
\begin{align}
\text{Re}(D_{jk})&=\frac{1}{2}C_{jk}\\
C_{jk}(\omega)&= \int_{-\infty}^\infty d \tau \; e^{i \omega \tau} \, \langle  \tilde B_j(\tau) \tilde B_k(0) \rangle 
\end{align}
The Markov approximation, which is used to derive the Bloch-Redfield equations, requires $C_{jk}(\omega)$ to be constant on the frequency scale of the quantum system, i.e.~``smoothness''. This means a fast decay of the \emph{temporal} correlation function, which is independent of the \emph{spatial} correlation length.

\section{Secular approximation of the Bloch-Redfield equations }
\label{app secular approximation of B-R}
The secular approximation can be used to simplify the Bloch-Redfield equations to the form of eq.~\eqref{Lindblad non-diagonal form}. 
First we separate the system operators $s_j$ into several operators by multiplication with projection operators onto the Hamiltonian eigenstates $\ket{\omega_n}$. 
\begin{align}
s_j(\epsilon)=\sum_{\omega_m-\omega_n=\epsilon} \ket{\omega_n}\bra{\omega_n} s_j \ket{\omega_m}\bra{\omega_m} \label{s_j of epsilon definition}
\end{align}
The sum extends over all Hamiltonian eigenvalues $\omega_n$ and $\omega_m$ with a fixed difference $\epsilon$. Assuming all $s_j$ to be Hermitian we find 
\begin{align}
s_j(-\epsilon)=s_j(\epsilon)^\dagger \label{s_j of minus eps is s_j dagger}
\end{align}
Using Eq.~\eqref{s_j of epsilon definition}, we write:
\begin{align}
s_j&=\sum_\epsilon s_j(\epsilon) \label{replacement 1}\\
q_{jk}&=V^\dagger \sum_\epsilon s_k(\epsilon) \frac{1}{2}C_{jk}(\epsilon) V\\
\hat q_{jk}&=V^\dagger \sum_\epsilon s_k(\epsilon) \frac{1}{2}C_{kj}(-\epsilon) V\label{replacement 3}
\end{align}
Replacing eq.~\eqref{replacement 1} to \eqref{replacement 3} in the Bloch-Redfield equations (eq.~\eqref{B-R equations}) and changing to the interaction picture:
\begin{align}
e^{i H_S t} s_j(\epsilon) e ^{-i H_S t} = e^{-i \epsilon t} s_j(\epsilon)
\end{align}
One then finds:
\begin{align*}
\dot{ \tilde{ \rho}} &= \sum_{jk\epsilon \epsilon'} e^{-i(\epsilon+\epsilon')t} C_{jk}(\epsilon) \left( -s_j(\epsilon') s_k(\epsilon) \tilde \rho + s_k(\epsilon) \tilde \rho s_j(\epsilon') \right)
\\ &+\sum_{jk\epsilon \epsilon'} e^{-i(\epsilon+\epsilon')t} 
C_{kj}(-\epsilon) \left( -\tilde \rho s_k(\epsilon) s_j(\epsilon') + s_j(\epsilon') \tilde \rho s_k(\epsilon) \right) 
\end{align*}
Since $\epsilon$ includes negative frequencies the only non-oscillating terms are for $\epsilon'=-\epsilon$. In the secular approximation, the oscillating terms which are neglected are fast oscillating on the time scale of the system dynamics. This time scale $\tau_S=1/|\epsilon-\epsilon'|$ for $\epsilon \neq \epsilon'$ must be much shorter than the decoherence time scale set by the terms $C_{jk}$ and the magnitudes of the $s_j$. Roughly this can be restated as $H_S \gg H_{int}$. Applying the secular approximation one finds:
\begin{align}
\dot{\tilde{\rho}} &= \sum_{jk\epsilon} C_{jk}(\epsilon) \left( -s_j(-\epsilon) s_k(\epsilon) \tilde \rho + s_k(\epsilon) \tilde \rho s_j(-\epsilon) \right) \nonumber \\
&+ \sum_{jk\epsilon}  C_{kj}(-\epsilon) \left( -\tilde \rho s_k(\epsilon) s_j(-\epsilon) + s_j(-\epsilon) \tilde \rho s_k(\epsilon) \right) 
\label{Zwischenergebnis1}
\end{align}
Since the sum over $\epsilon$ extends over all positive and negative energy differences we can replace $\epsilon \rightarrow -\epsilon$ in the second line of eq.~\eqref{Zwischenergebnis1}. Note that this makes the last two terms the Hermitian conjugate of the first two terms. Furthermore we swap the equivalent indices $j$ and $k$ in the last two terms. This together with eq.~\eqref{s_j of minus eps is s_j dagger} yields:
\begin{align}
\dot{\tilde{\rho}} &= \sum_\epsilon  \sum_{jk} \frac{1}{2} C_{jk}(\epsilon) \left( 2 s_k(\epsilon) \tilde \rho s_j(\epsilon)^\dagger - \left\{ s_j(\epsilon)^\dagger s_k(\epsilon) , \tilde \rho \right\} \right) \label{B-R with secular approx}
\end{align}

In several cases one finds only one frequency in the elements of $q_{jk}$ and $\hat q_{jk}$ in which case the summation over $\epsilon$ contains only one summand and can be omitted, simplifying eq.~\eqref{B-R with secular approx} (back in the Schr\"odinger picture) to: 
\begin{align}
\dot \rho = \frac{i}{\hbar} [\rho,H_s] + \frac{1}{\hbar^2} \sum_{j,k} \frac{1}{2} C_{jk} \left(2 s_k \rho s_j^\dagger - \{ s_j^\dagger s_k, \rho\} \right) \label{B-R with secular approx (no frequency dependence)}
\end{align}

For ease of notation, the form of eq.~(\eqref{B-R with secular approx (no frequency dependence)}) is used without loss of generality. For cases where the summation over $\epsilon$ in eq.~\eqref{B-R with secular approx} is relevant, each frequency component is added independently and all further results apply analogously. 
In practice the secular approximation for a particular system is best applied via the method given in appendix A1 of reference \cite{Jeske2012}. The general arguments above are analogous to the derivation of the Lindblad form on page 128 ff. in reference \cite{Breuerbook}.

For a system of two qubits 
$$H=\omega_q \sigma_z^{(q1)} + (\omega_q+\delta)\sigma_z^{(q2)} + H_c$$ 
the general coupling operator $H_c$ between the two (which consists of tensor products of all Pauli matrices acting on one qubit with all Pauli matrices on the other with respective coupling strengths) can be simplified via a rotating wave approximation (assuming $\omega_q$ much greater than $\delta$ and all coupling strengths) to only a transverse and a longitudinal component: 
\begin{align}
H_c &= v_\perp (\sigma_x^{(q1)} \sigma_x^{(q2)} + \sigma_y^{(q1)} \sigma_y^{(q2)})/2 + v_\parallel \sigma_z^{(q1)} \sigma_z^{(q2)} \nonumber\\
&= v_\perp (\sigma_+^{(q1)} \sigma_-^{(q2)} + \sigma_-^{(q1)} \sigma_+^{(q2)}) + v_\parallel \sigma_z^{(q1)} \sigma_z^{(q2)}
\end{align}

Regarding environmental coupling of a qubit the transverse component leads to energy exchange between qubit and environment and the longitudinal component simply imparts a phase shift on the qubit. The environmental coupling of the two qubits with transverse and longitudinal components is: 
$$H_{int}=\sigma_x^{(q1)} B_1 + \sigma_z^{(q2)} B_2 + \sigma_x^{(q2)}B_3 + \sigma_z^{(q2)} B_4$$
Applying the secular approximation (based on a large $\omega_q$) one finds that the two coupling types can be approximated as coupling to independent baths. For systems of several qubits this applies independently for each pair. Therefore no mixed couplings of transversal and longitudinal operators appear in the Bloch-Redfield equations in the secular approximation.

\section{Complex spectral functions}
\label{app complex spectral functions}
Strictly speaking the spectral function  $C_{jk}(\omega)$ is actually given by a one-sided Fourier transform (see eq.~\eqref{Laplace transform definition}) $D_{jk}(\omega)=C_{jk}(\omega)+i F_{jk}(\omega)$ which can be in general complex. Taking that into account eq.~\eqref{Zwischenergebnis1} becomes:
\begin{align}
\dot{\tilde{\rho}} &= \sum_{jk\epsilon} D_{jk}(\epsilon) \left( -s_j(-\epsilon) s_k(\epsilon) \tilde \rho + s_k(\epsilon) \tilde \rho s_j(-\epsilon) \right)\\
&+ \sum_{jk\epsilon} (D_{jk}(\epsilon))^* \left( -\tilde \rho s_k(-\epsilon) s_j(\epsilon) + s_j(\epsilon) \tilde \rho s_k(-\epsilon) \right)
\end{align}
Dividing the expression into its real and imaginary components, one finds additional terms to eq.~\eqref{B-R with secular approx}:
\begin{align}
\dot{\tilde{\rho}} &= \sum_\epsilon  \sum_{jk} \frac{1}{2} C_{jk}(\epsilon) \left( 2 s_k(\epsilon) \tilde \rho s_j(\epsilon)^\dagger - \left\{ s_j(\epsilon)^\dagger s_k(\epsilon) , \tilde \rho \right\} \right) \nonumber \\
&+\sum_\epsilon  \sum_{jk} i F_{jk}(\epsilon) \left( \rho s_j(\epsilon)^\dagger s_k(\epsilon) - s_j(\epsilon)^\dagger s_k(\epsilon) \rho  \right)
\end{align}
which (back in the Schr\"odinger picture) can be regarded as a correction to the system Hamiltonian, often referred to as Lamb shift terms \cite{Breuerbook}.
\begin{align}
&\dot \rho = \frac{i}{\hbar} [\rho,H_s + H_{cor}] \label{B-R with correction Hamiltonian} \\
&+ \frac{1}{\hbar^2} \sum_\epsilon  \sum_{jk} \frac{1}{2} C_{jk}(\epsilon) \left( 2 s_k(\epsilon) \tilde \rho s_j(\epsilon)^\dagger - \left\{ s_j(\epsilon)^\dagger s_k(\epsilon) , \tilde \rho \right\} \right) \nonumber \\
&H_{cor}=\frac{1}{\hbar} \sum_\epsilon  \sum_{jk} F_{jk}(\epsilon) s_j(\epsilon)^\dagger s_k(\epsilon)
\end{align}
Again for ease of notation the summation over $\epsilon$ is omitted in the main text.

\bibliography{publication}

\begin{thebibliography}{53}%
\makeatletter
\providecommand \@ifxundefined [1]{%
 \@ifx{#1\undefined}
}%
\providecommand \@ifnum [1]{%
 \ifnum #1\expandafter \@firstoftwo
 \else \expandafter \@secondoftwo
 \fi
}%
\providecommand \@ifx [1]{%
 \ifx #1\expandafter \@firstoftwo
 \else \expandafter \@secondoftwo
 \fi
}%
\providecommand \natexlab [1]{#1}%
\providecommand \enquote  [1]{``#1''}%
\providecommand \bibnamefont  [1]{#1}%
\providecommand \bibfnamefont [1]{#1}%
\providecommand \citenamefont [1]{#1}%
\providecommand \href@noop [0]{\@secondoftwo}%
\providecommand \href [0]{\begingroup \@sanitize@url \@href}%
\providecommand \@href[1]{\@@startlink{#1}\@@href}%
\providecommand \@@href[1]{\endgroup#1\@@endlink}%
\providecommand \@sanitize@url [0]{\catcode `\\12\catcode `\$12\catcode
  `\&12\catcode `\#12\catcode `\^12\catcode `\_12\catcode `\%12\relax}%
\providecommand \@@startlink[1]{}%
\providecommand \@@endlink[0]{}%
\providecommand \url  [0]{\begingroup\@sanitize@url \@url }%
\providecommand \@url [1]{\endgroup\@href {#1}{\urlprefix }}%
\providecommand \urlprefix  [0]{URL }%
\providecommand \Eprint [0]{\href }%
\providecommand \doibase [0]{http://dx.doi.org/}%
\providecommand \selectlanguage [0]{\@gobble}%
\providecommand \bibinfo  [0]{\@secondoftwo}%
\providecommand \bibfield  [0]{\@secondoftwo}%
\providecommand \translation [1]{[#1]}%
\providecommand \BibitemOpen [0]{}%
\providecommand \bibitemStop [0]{}%
\providecommand \bibitemNoStop [0]{.\EOS\space}%
\providecommand \EOS [0]{\spacefactor3000\relax}%
\providecommand \BibitemShut  [1]{\csname bibitem#1\endcsname}%
\let\auto@bib@innerbib\@empty
\bibitem [{\citenamefont {Breuer}\ and\ \citenamefont
  {Petruccione}(2003)}]{Breuerbook}%
  \BibitemOpen
  \bibfield  {author} {\bibinfo {author} {\bibfnamefont {H.-P.}\ \bibnamefont
  {Breuer}}\ and\ \bibinfo {author} {\bibfnamefont {F.}~\bibnamefont
  {Petruccione}},\ }\href@noop {} {\emph {\bibinfo {title} {The theory of open
  quantum systems}}}\ (\bibinfo  {publisher} {Oxford University Press},\
  \bibinfo {year} {2003})\BibitemShut {NoStop}%
\bibitem [{\citenamefont {Terhal}\ and\ \citenamefont
  {Burkard}(2005)}]{Terhal2005}%
  \BibitemOpen
  \bibfield  {author} {\bibinfo {author} {\bibfnamefont {B.~M.}\ \bibnamefont
  {Terhal}}\ and\ \bibinfo {author} {\bibfnamefont {G.}~\bibnamefont
  {Burkard}},\ }\href {\doibase 10.1103/PhysRevA.71.012336} {\bibfield
  {journal} {\bibinfo  {journal} {Phys. Rev. A}\ }\textbf {\bibinfo {volume}
  {71}},\ \bibinfo {pages} {012336} (\bibinfo {year} {2005})}\BibitemShut
  {NoStop}%
\bibitem [{\citenamefont {Doll}\ \emph {et~al.}(2007)\citenamefont {Doll},
  \citenamefont {Wubs}, \citenamefont {H\"anggi},\ and\ \citenamefont
  {Kohler}}]{Doll2007}%
  \BibitemOpen
  \bibfield  {author} {\bibinfo {author} {\bibfnamefont {R.}~\bibnamefont
  {Doll}}, \bibinfo {author} {\bibfnamefont {M.}~\bibnamefont {Wubs}}, \bibinfo
  {author} {\bibfnamefont {P.}~\bibnamefont {H\"anggi}}, \ and\ \bibinfo
  {author} {\bibfnamefont {S.}~\bibnamefont {Kohler}},\ }\href {\doibase
  10.1103/PhysRevB.76.045317} {\bibfield  {journal} {\bibinfo  {journal} {Phys.
  Rev. B}\ }\textbf {\bibinfo {volume} {76}},\ \bibinfo {pages} {045317}
  (\bibinfo {year} {2007})}\BibitemShut {NoStop}%
\bibitem [{\citenamefont {Contreras-Pulido}\ and\ \citenamefont
  {Aguado}(2008)}]{Contreras2008}%
  \BibitemOpen
  \bibfield  {author} {\bibinfo {author} {\bibfnamefont {L.~D.}\ \bibnamefont
  {Contreras-Pulido}}\ and\ \bibinfo {author} {\bibfnamefont {R.}~\bibnamefont
  {Aguado}},\ }\href {\doibase 10.1103/PhysRevB.77.155420} {\bibfield
  {journal} {\bibinfo  {journal} {Phys. Rev. B}\ }\textbf {\bibinfo {volume}
  {77}},\ \bibinfo {pages} {155420} (\bibinfo {year} {2008})}\BibitemShut
  {NoStop}%
\bibitem [{\citenamefont {McCutcheon}\ \emph {et~al.}(2009)\citenamefont
  {McCutcheon}, \citenamefont {Nazir}, \citenamefont {Bose},\ and\
  \citenamefont {Fisher}}]{McCutcheon2009}%
  \BibitemOpen
  \bibfield  {author} {\bibinfo {author} {\bibfnamefont {D.~P.~S.}\
  \bibnamefont {McCutcheon}}, \bibinfo {author} {\bibfnamefont
  {A.}~\bibnamefont {Nazir}}, \bibinfo {author} {\bibfnamefont
  {S.}~\bibnamefont {Bose}}, \ and\ \bibinfo {author} {\bibfnamefont {A.~J.}\
  \bibnamefont {Fisher}},\ }\href {\doibase 10.1103/PhysRevA.80.022337}
  {\bibfield  {journal} {\bibinfo  {journal} {Phys. Rev. A}\ }\textbf {\bibinfo
  {volume} {80}},\ \bibinfo {pages} {022337} (\bibinfo {year}
  {2009})}\BibitemShut {NoStop}%
\bibitem [{\citenamefont {Jeske}\ \emph {et~al.}(2012)\citenamefont {Jeske},
  \citenamefont {Cole}, \citenamefont {M\"uller}, \citenamefont {Marthaler},\
  and\ \citenamefont {Sch\"on}}]{Jeske2012}%
  \BibitemOpen
  \bibfield  {author} {\bibinfo {author} {\bibfnamefont {J.}~\bibnamefont
  {Jeske}}, \bibinfo {author} {\bibfnamefont {J.~H.}\ \bibnamefont {Cole}},
  \bibinfo {author} {\bibfnamefont {C.}~\bibnamefont {M\"uller}}, \bibinfo
  {author} {\bibfnamefont {M.}~\bibnamefont {Marthaler}}, \ and\ \bibinfo
  {author} {\bibfnamefont {G.}~\bibnamefont {Sch\"on}},\ }\href
  {http://stacks.iop.org/1367-2630/14/i=2/a=023013} {\bibfield  {journal}
  {\bibinfo  {journal} {New Journal of Physics}\ }\textbf {\bibinfo {volume}
  {14}},\ \bibinfo {pages} {023013} (\bibinfo {year} {2012})}\BibitemShut
  {NoStop}%
\bibitem [{\citenamefont {Dicke}(1954)}]{Dicke1954}%
  \BibitemOpen
  \bibfield  {author} {\bibinfo {author} {\bibfnamefont {R.~H.}\ \bibnamefont
  {Dicke}},\ }\href {\doibase 10.1103/PhysRev.93.99} {\bibfield  {journal}
  {\bibinfo  {journal} {Phys. Rev.}\ }\textbf {\bibinfo {volume} {93}},\
  \bibinfo {pages} {99} (\bibinfo {year} {1954})}\BibitemShut {NoStop}%
\bibitem [{\citenamefont {Crubellier}\ \emph {et~al.}(1985)\citenamefont
  {Crubellier}, \citenamefont {Liberman}, \citenamefont {Pavolini},\ and\
  \citenamefont {Pillet}}]{Crubellier1985}%
  \BibitemOpen
  \bibfield  {author} {\bibinfo {author} {\bibfnamefont {A.}~\bibnamefont
  {Crubellier}}, \bibinfo {author} {\bibfnamefont {S.}~\bibnamefont
  {Liberman}}, \bibinfo {author} {\bibfnamefont {D.}~\bibnamefont {Pavolini}},
  \ and\ \bibinfo {author} {\bibfnamefont {P.}~\bibnamefont {Pillet}},\ }\href
  {http://stacks.iop.org/0022-3700/18/i=18/a=022} {\bibfield  {journal}
  {\bibinfo  {journal} {Journal of Physics B}\ }\textbf {\bibinfo {volume}
  {18}},\ \bibinfo {pages} {3811} (\bibinfo {year} {1985})}\BibitemShut
  {NoStop}%
\bibitem [{\citenamefont {Lidar}\ \emph {et~al.}(1998)\citenamefont {Lidar},
  \citenamefont {Chuang},\ and\ \citenamefont {Whaley}}]{Lidar1998}%
  \BibitemOpen
  \bibfield  {author} {\bibinfo {author} {\bibfnamefont {D.~A.}\ \bibnamefont
  {Lidar}}, \bibinfo {author} {\bibfnamefont {I.~L.}\ \bibnamefont {Chuang}}, \
  and\ \bibinfo {author} {\bibfnamefont {K.~B.}\ \bibnamefont {Whaley}},\
  }\href {\doibase 10.1103/PhysRevLett.81.2594} {\bibfield  {journal} {\bibinfo
   {journal} {Phys Rev Lett}\ }\textbf {\bibinfo {volume} {81}},\ \bibinfo
  {pages} {2594} (\bibinfo {year} {1998})}\BibitemShut {NoStop}%
\bibitem [{\citenamefont {Lidar}\ \emph
  {et~al.}(2001{\natexlab{a}})\citenamefont {Lidar}, \citenamefont {Bacon},
  \citenamefont {Kempe},\ and\ \citenamefont {Whaley}}]{Lidar2001}%
  \BibitemOpen
  \bibfield  {author} {\bibinfo {author} {\bibfnamefont {D.~A.}\ \bibnamefont
  {Lidar}}, \bibinfo {author} {\bibfnamefont {D.}~\bibnamefont {Bacon}},
  \bibinfo {author} {\bibfnamefont {J.}~\bibnamefont {Kempe}}, \ and\ \bibinfo
  {author} {\bibfnamefont {K.~B.}\ \bibnamefont {Whaley}},\ }\href {\doibase
  10.1103/PhysRevA.63.022306} {\bibfield  {journal} {\bibinfo  {journal} {Phys.
  Rev. A}\ }\textbf {\bibinfo {volume} {63}},\ \bibinfo {pages} {022306}
  (\bibinfo {year} {2001}{\natexlab{a}})}\BibitemShut {NoStop}%
\bibitem [{\citenamefont {Lidar}\ \emph
  {et~al.}(2001{\natexlab{b}})\citenamefont {Lidar}, \citenamefont {Bacon},
  \citenamefont {Kempe},\ and\ \citenamefont {Whaley}}]{Lidar2001_II}%
  \BibitemOpen
  \bibfield  {author} {\bibinfo {author} {\bibfnamefont {D.~A.}\ \bibnamefont
  {Lidar}}, \bibinfo {author} {\bibfnamefont {D.}~\bibnamefont {Bacon}},
  \bibinfo {author} {\bibfnamefont {J.}~\bibnamefont {Kempe}}, \ and\ \bibinfo
  {author} {\bibfnamefont {K.~B.}\ \bibnamefont {Whaley}},\ }\href {\doibase
  10.1103/PhysRevA.63.022307} {\bibfield  {journal} {\bibinfo  {journal} {Phys.
  Rev. A}\ }\textbf {\bibinfo {volume} {63}},\ \bibinfo {pages} {022307}
  (\bibinfo {year} {2001}{\natexlab{b}})}\BibitemShut {NoStop}%
\bibitem [{\citenamefont {Lidar}\ and\ \citenamefont
  {Whaley}(2003)}]{Lidar2003}%
  \BibitemOpen
  \bibfield  {author} {\bibinfo {author} {\bibfnamefont {D.~A.}\ \bibnamefont
  {Lidar}}\ and\ \bibinfo {author} {\bibfnamefont {K.~B.}\ \bibnamefont
  {Whaley}},\ }in\ \href@noop {} {\emph {\bibinfo {booktitle} {Irreversible
  Quantum Dynamics}}},\ \bibinfo {series} {Lecture Notes in Physics}, Vol.\
  \bibinfo {volume} {622},\ \bibinfo {editor} {edited by\ \bibinfo {editor}
  {\bibfnamefont {F.}~\bibnamefont {Benatti}}\ and\ \bibinfo {editor}
  {\bibfnamefont {R.}~\bibnamefont {Floreanini}}}\ (\bibinfo  {publisher}
  {Springer Berlin},\ \bibinfo {year} {2003})\ pp.\ \bibinfo {pages}
  {83--120}\BibitemShut {NoStop}%
\bibitem [{\citenamefont {Blume-Kohout}\ \emph {et~al.}(2008)\citenamefont
  {Blume-Kohout}, \citenamefont {Ng}, \citenamefont {Poulin},\ and\
  \citenamefont {Viola}}]{Blume-Kohout2008}%
  \BibitemOpen
  \bibfield  {author} {\bibinfo {author} {\bibfnamefont {R.}~\bibnamefont
  {Blume-Kohout}}, \bibinfo {author} {\bibfnamefont {H.~K.}\ \bibnamefont
  {Ng}}, \bibinfo {author} {\bibfnamefont {D.}~\bibnamefont {Poulin}}, \ and\
  \bibinfo {author} {\bibfnamefont {L.}~\bibnamefont {Viola}},\ }\href
  {\doibase 10.1103/PhysRevLett.100.030501} {\bibfield  {journal} {\bibinfo
  {journal} {Phys. Rev. Lett.}\ }\textbf {\bibinfo {volume} {100}},\ \bibinfo
  {pages} {030501} (\bibinfo {year} {2008})}\BibitemShut {NoStop}%
\bibitem [{\citenamefont {Doll}\ \emph {et~al.}(2006)\citenamefont {Doll},
  \citenamefont {Wubs}, \citenamefont {H\"anggi},\ and\ \citenamefont
  {Kohler}}]{Doll2006}%
  \BibitemOpen
  \bibfield  {author} {\bibinfo {author} {\bibfnamefont {R.}~\bibnamefont
  {Doll}}, \bibinfo {author} {\bibfnamefont {M.}~\bibnamefont {Wubs}}, \bibinfo
  {author} {\bibfnamefont {P.}~\bibnamefont {H\"anggi}}, \ and\ \bibinfo
  {author} {\bibfnamefont {S.}~\bibnamefont {Kohler}},\ }\href
  {http://stacks.iop.org/0295-5075/76/i=4/a=547} {\bibfield  {journal}
  {\bibinfo  {journal} {EPL (Europhysics Letters)}\ }\textbf {\bibinfo {volume}
  {76}},\ \bibinfo {pages} {547} (\bibinfo {year} {2006})}\BibitemShut
  {NoStop}%
\bibitem [{\citenamefont {Palma}\ \emph {et~al.}(1996)\citenamefont {Palma},
  \citenamefont {Suominen},\ and\ \citenamefont {Ekert}}]{Palma1996}%
  \BibitemOpen
  \bibfield  {author} {\bibinfo {author} {\bibfnamefont {G.}~\bibnamefont
  {Palma}}, \bibinfo {author} {\bibfnamefont {K.}~\bibnamefont {Suominen}}, \
  and\ \bibinfo {author} {\bibfnamefont {A.}~\bibnamefont {Ekert}},\
  }\href@noop {} {\bibfield  {journal} {\bibinfo  {journal} {Proceedings of the
  Royal Society a-Mathematical Physical and Engineering Sciences}\ }\textbf
  {\bibinfo {volume} {452}},\ \bibinfo {pages} {567} (\bibinfo {year}
  {1996})}\BibitemShut {NoStop}%
\bibitem [{\citenamefont {Reina}\ \emph {et~al.}(2002)\citenamefont {Reina},
  \citenamefont {Quiroga},\ and\ \citenamefont {Johnson}}]{Reina2002}%
  \BibitemOpen
  \bibfield  {author} {\bibinfo {author} {\bibfnamefont {J.}~\bibnamefont
  {Reina}}, \bibinfo {author} {\bibfnamefont {L.}~\bibnamefont {Quiroga}}, \
  and\ \bibinfo {author} {\bibfnamefont {N.}~\bibnamefont {Johnson}},\
  }\href@noop {} {\bibfield  {journal} {\bibinfo  {journal} {Physical Review
  A}\ }\textbf {\bibinfo {volume} {65}},\ \bibinfo {pages} {032326} (\bibinfo
  {year} {2002})}\BibitemShut {NoStop}%
\bibitem [{\citenamefont {Duan}\ and\ \citenamefont {Guo}(1998)}]{Duan1998}%
  \BibitemOpen
  \bibfield  {author} {\bibinfo {author} {\bibfnamefont {L.-M.}\ \bibnamefont
  {Duan}}\ and\ \bibinfo {author} {\bibfnamefont {G.-C.}\ \bibnamefont {Guo}},\
  }\href {\doibase 10.1103/PhysRevA.57.737} {\bibfield  {journal} {\bibinfo
  {journal} {Phys. Rev. A}\ }\textbf {\bibinfo {volume} {57}},\ \bibinfo
  {pages} {737} (\bibinfo {year} {1998})}\BibitemShut {NoStop}%
\bibitem [{\citenamefont {Zanardi}(1998)}]{Zanardi1998}%
  \BibitemOpen
  \bibfield  {author} {\bibinfo {author} {\bibfnamefont {P.}~\bibnamefont
  {Zanardi}},\ }\href {\doibase 10.1103/PhysRevA.57.3276} {\bibfield  {journal}
  {\bibinfo  {journal} {Physical Review A}\ }\textbf {\bibinfo {volume} {57}},\
  \bibinfo {pages} {3276} (\bibinfo {year} {1998})}\BibitemShut {NoStop}%
\bibitem [{\citenamefont {Alicki}\ \emph {et~al.}(2002)\citenamefont {Alicki},
  \citenamefont {Horodecki}, \citenamefont {Horodecki},\ and\ \citenamefont
  {Horodecki}}]{Alicki2002}%
  \BibitemOpen
  \bibfield  {author} {\bibinfo {author} {\bibfnamefont {R.}~\bibnamefont
  {Alicki}}, \bibinfo {author} {\bibfnamefont {M.}~\bibnamefont {Horodecki}},
  \bibinfo {author} {\bibfnamefont {P.}~\bibnamefont {Horodecki}}, \ and\
  \bibinfo {author} {\bibfnamefont {R.}~\bibnamefont {Horodecki}},\ }\href
  {\doibase 10.1103/PhysRevA.65.062101} {\bibfield  {journal} {\bibinfo
  {journal} {Physical Review A}\ }\textbf {\bibinfo {volume} {65}},\ \bibinfo
  {pages} {062101} (\bibinfo {year} {2002})}\BibitemShut {NoStop}%
\bibitem [{\citenamefont {Averin}\ and\ \citenamefont
  {Fazio}(2003)}]{Averin2003}%
  \BibitemOpen
  \bibfield  {author} {\bibinfo {author} {\bibfnamefont {D.~V.}\ \bibnamefont
  {Averin}}\ and\ \bibinfo {author} {\bibfnamefont {R.}~\bibnamefont {Fazio}},\
  }\href@noop {} {\bibfield  {journal} {\bibinfo  {journal} {JETP Lett.}\
  }\textbf {\bibinfo {volume} {78}},\ \bibinfo {pages} {664} (\bibinfo {year}
  {2003})}\BibitemShut {NoStop}%
\bibitem [{\citenamefont {Clemens}\ \emph {et~al.}(2004)\citenamefont
  {Clemens}, \citenamefont {Siddiqui},\ and\ \citenamefont
  {Gea-Banacloche}}]{Clemens2004}%
  \BibitemOpen
  \bibfield  {author} {\bibinfo {author} {\bibfnamefont {J.~P.}\ \bibnamefont
  {Clemens}}, \bibinfo {author} {\bibfnamefont {S.}~\bibnamefont {Siddiqui}}, \
  and\ \bibinfo {author} {\bibfnamefont {J.}~\bibnamefont {Gea-Banacloche}},\
  }\href {\doibase 10.1103/PhysRevA.69.062313} {\bibfield  {journal} {\bibinfo
  {journal} {Phys. Rev. A}\ }\textbf {\bibinfo {volume} {69}},\ \bibinfo
  {pages} {062313} (\bibinfo {year} {2004})}\BibitemShut {NoStop}%
\bibitem [{\citenamefont {Aharonov}\ \emph {et~al.}(2006)\citenamefont
  {Aharonov}, \citenamefont {Kitaev},\ and\ \citenamefont
  {Preskill}}]{Aharonov2006}%
  \BibitemOpen
  \bibfield  {author} {\bibinfo {author} {\bibfnamefont {D.}~\bibnamefont
  {Aharonov}}, \bibinfo {author} {\bibfnamefont {A.}~\bibnamefont {Kitaev}}, \
  and\ \bibinfo {author} {\bibfnamefont {J.}~\bibnamefont {Preskill}},\ }\href
  {\doibase 10.1103/PhysRevLett.96.050504} {\bibfield  {journal} {\bibinfo
  {journal} {Phys. Rev. Lett.}\ }\textbf {\bibinfo {volume} {96}},\ \bibinfo
  {pages} {050504} (\bibinfo {year} {2006})}\BibitemShut {NoStop}%
\bibitem [{\citenamefont {Aliferis}\ \emph {et~al.}(2006)\citenamefont
  {Aliferis}, \citenamefont {Gottesman},\ and\ \citenamefont
  {Preskill}}]{Aliferis2006}%
  \BibitemOpen
  \bibfield  {author} {\bibinfo {author} {\bibfnamefont {P.}~\bibnamefont
  {Aliferis}}, \bibinfo {author} {\bibfnamefont {D.}~\bibnamefont {Gottesman}},
  \ and\ \bibinfo {author} {\bibfnamefont {J.}~\bibnamefont {Preskill}},\
  }\href {\doibase http://www.rintonpress.com/xqic6/qic-6-2/097-165.pdf}
  {\bibfield  {journal} {\bibinfo  {journal} {Quantum Inf. Comput.}\ }\textbf
  {\bibinfo {volume} {6}},\ \bibinfo {pages} {97} (\bibinfo {year}
  {2006})}\BibitemShut {NoStop}%
\bibitem [{\citenamefont {Novais}\ and\ \citenamefont
  {Baranger}(2006)}]{Novais2006}%
  \BibitemOpen
  \bibfield  {author} {\bibinfo {author} {\bibfnamefont {E.}~\bibnamefont
  {Novais}}\ and\ \bibinfo {author} {\bibfnamefont {H.~U.}\ \bibnamefont
  {Baranger}},\ }\href {\doibase 10.1103/PhysRevLett.97.040501} {\bibfield
  {journal} {\bibinfo  {journal} {Phys. Rev. Lett.}\ }\textbf {\bibinfo
  {volume} {97}},\ \bibinfo {pages} {040501} (\bibinfo {year}
  {2006})}\BibitemShut {NoStop}%
\bibitem [{\citenamefont {Caruso}\ \emph {et~al.}(2009)\citenamefont {Caruso},
  \citenamefont {Chin}, \citenamefont {Datta}, \citenamefont {Huelga},\ and\
  \citenamefont {Plenio}}]{Caruso2009}%
  \BibitemOpen
  \bibfield  {author} {\bibinfo {author} {\bibfnamefont {F.}~\bibnamefont
  {Caruso}}, \bibinfo {author} {\bibfnamefont {A.~W.}\ \bibnamefont {Chin}},
  \bibinfo {author} {\bibfnamefont {A.}~\bibnamefont {Datta}}, \bibinfo
  {author} {\bibfnamefont {S.~F.}\ \bibnamefont {Huelga}}, \ and\ \bibinfo
  {author} {\bibfnamefont {M.~B.}\ \bibnamefont {Plenio}},\ }\href {\doibase
  10.1063/1.3223548} {\bibfield  {journal} {\bibinfo  {journal} {Journal of
  Chemical Physics}\ }\textbf {\bibinfo {volume} {131}},\ \bibinfo {eid}
  {105106} (\bibinfo {year} {2009})}\BibitemShut {NoStop}%
\bibitem [{\citenamefont {Scholes}\ \emph {et~al.}(2012)\citenamefont
  {Scholes}, \citenamefont {Mirkovic}, \citenamefont {Turner}, \citenamefont
  {Fassioli},\ and\ \citenamefont {Buchleitner}}]{Scholes2012}%
  \BibitemOpen
  \bibfield  {author} {\bibinfo {author} {\bibfnamefont {G.~D.}\ \bibnamefont
  {Scholes}}, \bibinfo {author} {\bibfnamefont {T.}~\bibnamefont {Mirkovic}},
  \bibinfo {author} {\bibfnamefont {D.~B.}\ \bibnamefont {Turner}}, \bibinfo
  {author} {\bibfnamefont {F.}~\bibnamefont {Fassioli}}, \ and\ \bibinfo
  {author} {\bibfnamefont {A.}~\bibnamefont {Buchleitner}},\ }\href {\doibase
  10.1039/C2EE23013E} {\bibfield  {journal} {\bibinfo  {journal} {Energy
  Environ. Sci.}\ }\textbf {\bibinfo {volume} {5}},\ \bibinfo {pages} {9374}
  (\bibinfo {year} {2012})}\BibitemShut {NoStop}%
\bibitem [{\citenamefont {Yang}\ and\ \citenamefont
  {Fleming}(2002)}]{Yang2002}%
  \BibitemOpen
  \bibfield  {author} {\bibinfo {author} {\bibfnamefont {M.}~\bibnamefont
  {Yang}}\ and\ \bibinfo {author} {\bibfnamefont {G.~R.}\ \bibnamefont
  {Fleming}},\ }\href {\doibase 10.1016/S0301-0104(02)00604-3} {\bibfield
  {journal} {\bibinfo  {journal} {Chemical Physics}\ }\textbf {\bibinfo
  {volume} {282}},\ \bibinfo {pages} {163 } (\bibinfo {year}
  {2002})}\BibitemShut {NoStop}%
\bibitem [{\citenamefont {Palmieri}\ \emph {et~al.}(2009)\citenamefont
  {Palmieri}, \citenamefont {Abramavicius},\ and\ \citenamefont
  {Mukamel}}]{Palmieri2009}%
  \BibitemOpen
  \bibfield  {author} {\bibinfo {author} {\bibfnamefont {B.}~\bibnamefont
  {Palmieri}}, \bibinfo {author} {\bibfnamefont {D.}~\bibnamefont
  {Abramavicius}}, \ and\ \bibinfo {author} {\bibfnamefont {S.}~\bibnamefont
  {Mukamel}},\ }\href@noop {} {\bibfield  {journal} {\bibinfo  {journal} {J.
  Chem. Phys.}\ }\textbf {\bibinfo {volume} {130}},\ \bibinfo {pages} {204512}
  (\bibinfo {year} {2009})}\BibitemShut {NoStop}%
\bibitem [{\citenamefont {Ringsmuth}\ \emph {et~al.}(2012)\citenamefont
  {Ringsmuth}, \citenamefont {Milburn},\ and\ \citenamefont
  {Stace}}]{Ringsmuth2012}%
  \BibitemOpen
  \bibfield  {author} {\bibinfo {author} {\bibfnamefont {A.~K.}\ \bibnamefont
  {Ringsmuth}}, \bibinfo {author} {\bibfnamefont {G.~J.}\ \bibnamefont
  {Milburn}}, \ and\ \bibinfo {author} {\bibfnamefont {T.~M.}\ \bibnamefont
  {Stace}},\ }\href {\doibase 10.1038/nphys2332} {\bibfield  {journal}
  {\bibinfo  {journal} {Nature Physics}\ }\textbf {\bibinfo {volume} {8}},\
  \bibinfo {pages} {562} (\bibinfo {year} {2012})}\BibitemShut {NoStop}%
\bibitem [{\citenamefont {Safavi-Naini}\ \emph {et~al.}(2011)\citenamefont
  {Safavi-Naini}, \citenamefont {Rabl}, \citenamefont {Weck},\ and\
  \citenamefont {Sadeghpour}}]{Safavi-Naini2011}%
  \BibitemOpen
  \bibfield  {author} {\bibinfo {author} {\bibfnamefont {A.}~\bibnamefont
  {Safavi-Naini}}, \bibinfo {author} {\bibfnamefont {P.}~\bibnamefont {Rabl}},
  \bibinfo {author} {\bibfnamefont {P.~F.}\ \bibnamefont {Weck}}, \ and\
  \bibinfo {author} {\bibfnamefont {H.~R.}\ \bibnamefont {Sadeghpour}},\ }\href
  {\doibase 10.1103/PhysRevA.84.023412} {\bibfield  {journal} {\bibinfo
  {journal} {Phys. Rev. A}\ }\textbf {\bibinfo {volume} {84}},\ \bibinfo
  {pages} {023412} (\bibinfo {year} {2011})}\BibitemShut {NoStop}%
\bibitem [{\citenamefont {Monz}\ \emph {et~al.}(2011)\citenamefont {Monz},
  \citenamefont {Schindler}, \citenamefont {Barreiro}, \citenamefont {Chwalla},
  \citenamefont {Nigg}, \citenamefont {Coish}, \citenamefont {Harlander},
  \citenamefont {H\"ansel}, \citenamefont {Hennrich},\ and\ \citenamefont
  {Blatt}}]{Monz2011}%
  \BibitemOpen
  \bibfield  {author} {\bibinfo {author} {\bibfnamefont {T.}~\bibnamefont
  {Monz}}, \bibinfo {author} {\bibfnamefont {P.}~\bibnamefont {Schindler}},
  \bibinfo {author} {\bibfnamefont {J.~T.}\ \bibnamefont {Barreiro}}, \bibinfo
  {author} {\bibfnamefont {M.}~\bibnamefont {Chwalla}}, \bibinfo {author}
  {\bibfnamefont {D.}~\bibnamefont {Nigg}}, \bibinfo {author} {\bibfnamefont
  {W.~A.}\ \bibnamefont {Coish}}, \bibinfo {author} {\bibfnamefont
  {M.}~\bibnamefont {Harlander}}, \bibinfo {author} {\bibfnamefont
  {W.}~\bibnamefont {H\"ansel}}, \bibinfo {author} {\bibfnamefont
  {M.}~\bibnamefont {Hennrich}}, \ and\ \bibinfo {author} {\bibfnamefont
  {R.}~\bibnamefont {Blatt}},\ }\href {\doibase 10.1103/PhysRevLett.106.130506}
  {\bibfield  {journal} {\bibinfo  {journal} {Phys. Rev. Lett.}\ }\textbf
  {\bibinfo {volume} {106}},\ \bibinfo {pages} {130506} (\bibinfo {year}
  {2011})}\BibitemShut {NoStop}%
\bibitem [{\citenamefont {Deslauriers}\ \emph {et~al.}(2006)\citenamefont
  {Deslauriers}, \citenamefont {Olmschenk}, \citenamefont {Stick},
  \citenamefont {Hensinger}, \citenamefont {Sterk},\ and\ \citenamefont
  {Monroe}}]{Deslauriers2006}%
  \BibitemOpen
  \bibfield  {author} {\bibinfo {author} {\bibfnamefont {L.}~\bibnamefont
  {Deslauriers}}, \bibinfo {author} {\bibfnamefont {S.}~\bibnamefont
  {Olmschenk}}, \bibinfo {author} {\bibfnamefont {D.}~\bibnamefont {Stick}},
  \bibinfo {author} {\bibfnamefont {W.~K.}\ \bibnamefont {Hensinger}}, \bibinfo
  {author} {\bibfnamefont {J.}~\bibnamefont {Sterk}}, \ and\ \bibinfo {author}
  {\bibfnamefont {C.}~\bibnamefont {Monroe}},\ }\href {\doibase
  10.1103/PhysRevLett.97.103007} {\bibfield  {journal} {\bibinfo  {journal}
  {Phys. Rev. Lett.}\ }\textbf {\bibinfo {volume} {97}},\ \bibinfo {pages}
  {103007} (\bibinfo {year} {2006})}\BibitemShut {NoStop}%
\bibitem [{\citenamefont {Schroll}\ \emph {et~al.}(2003)\citenamefont
  {Schroll}, \citenamefont {Belzig},\ and\ \citenamefont
  {Bruder}}]{Schroll2003}%
  \BibitemOpen
  \bibfield  {author} {\bibinfo {author} {\bibfnamefont {C.}~\bibnamefont
  {Schroll}}, \bibinfo {author} {\bibfnamefont {W.}~\bibnamefont {Belzig}}, \
  and\ \bibinfo {author} {\bibfnamefont {C.}~\bibnamefont {Bruder}},\ }\href
  {\doibase 10.1103/PhysRevA.68.043618} {\bibfield  {journal} {\bibinfo
  {journal} {Phys. Rev. A}\ }\textbf {\bibinfo {volume} {68}},\ \bibinfo
  {pages} {043618} (\bibinfo {year} {2003})}\BibitemShut {NoStop}%
\bibitem [{\citenamefont {Lindblad}(1976)}]{Lindblad1976}%
  \BibitemOpen
  \bibfield  {author} {\bibinfo {author} {\bibfnamefont {G.}~\bibnamefont
  {Lindblad}},\ }\href {http://dx.doi.org/10.1007/BF01608499} {\bibfield
  {journal} {\bibinfo  {journal} {Communications in Mathematical Physics}\
  }\textbf {\bibinfo {volume} {48}},\ \bibinfo {pages} {119} (\bibinfo {year}
  {1976})},\ \bibinfo {note} {10.1007/BF01608499}\BibitemShut {NoStop}%
\bibitem [{\citenamefont {Gorini}\ \emph {et~al.}(1976)\citenamefont {Gorini},
  \citenamefont {Kossakowski},\ and\ \citenamefont {Sudarshan}}]{Gorini1976}%
  \BibitemOpen
  \bibfield  {author} {\bibinfo {author} {\bibfnamefont {V.}~\bibnamefont
  {Gorini}}, \bibinfo {author} {\bibfnamefont {A.}~\bibnamefont {Kossakowski}},
  \ and\ \bibinfo {author} {\bibfnamefont {E.~C.~G.}\ \bibnamefont
  {Sudarshan}},\ }\href@noop {} {\bibfield  {journal} {\bibinfo  {journal}
  {Journal of Mathematical Physics}\ }\textbf {\bibinfo {volume} {17}},\
  \bibinfo {pages} {821} (\bibinfo {year} {1976})}\BibitemShut {NoStop}%
\bibitem [{\citenamefont {Carmichael}(1999)}]{Carmichaelbook}%
  \BibitemOpen
  \bibfield  {author} {\bibinfo {author} {\bibfnamefont {H.~J.}\ \bibnamefont
  {Carmichael}},\ }\href@noop {} {\emph {\bibinfo {title} {Statistical methods
  in quantum optics 1}}}\ (\bibinfo  {publisher} {Springer},\ \bibinfo {year}
  {1999})\BibitemShut {NoStop}%
\bibitem [{\citenamefont {Van~Hove}(1954)}]{vanHove1954}%
  \BibitemOpen
  \bibfield  {author} {\bibinfo {author} {\bibfnamefont {L.}~\bibnamefont
  {Van~Hove}},\ }\href {\doibase 10.1103/PhysRev.95.249} {\bibfield  {journal}
  {\bibinfo  {journal} {Physical Review}\ }\textbf {\bibinfo {volume} {95}},\
  \bibinfo {pages} {249} (\bibinfo {year} {1954})}\BibitemShut {NoStop}%
\bibitem [{\citenamefont {Nolting}\ and\ \citenamefont
  {Ramakanth}(2009)}]{Noltingmagnetism}%
  \BibitemOpen
  \bibfield  {author} {\bibinfo {author} {\bibfnamefont {W.}~\bibnamefont
  {Nolting}}\ and\ \bibinfo {author} {\bibfnamefont {A.}~\bibnamefont
  {Ramakanth}},\ }\href {\doibase 10.1007/978-3-540-85416-6} {\emph {\bibinfo
  {title} {Quantum Theory of Magnetism}}}\ (\bibinfo  {publisher} {Springer
  Berlin Heidelberg},\ \bibinfo {year} {2009})\BibitemShut {NoStop}%
\bibitem [{\citenamefont {Glauber}(1963)}]{Glauber1963}%
  \BibitemOpen
  \bibfield  {author} {\bibinfo {author} {\bibfnamefont {R.~J.}\ \bibnamefont
  {Glauber}},\ }\href@noop {} {\bibfield  {journal} {\bibinfo  {journal}
  {Journal of Mathematical Physics}\ }\textbf {\bibinfo {volume} {4}},\
  \bibinfo {pages} {294} (\bibinfo {year} {1963})}\BibitemShut {NoStop}%
\bibitem [{\citenamefont {Kawasaki}(1972)}]{Kawasaki1972}%
  \BibitemOpen
  \bibfield  {author} {\bibinfo {author} {\bibfnamefont {K.}~\bibnamefont
  {Kawasaki}},\ }in\ \href@noop {} {\emph {\bibinfo {booktitle} {Phase
  transitions and critical phenomena}}},\ \bibinfo {editor} {edited by\
  \bibinfo {editor} {\bibfnamefont {C.}~\bibnamefont {Domb}}\ and\ \bibinfo
  {editor} {\bibfnamefont {M.}~\bibnamefont {Green}}}\ (\bibinfo  {publisher}
  {Academic Press London},\ \bibinfo {year} {1972})\BibitemShut {NoStop}%
\bibitem [{\citenamefont {Leggett}\ \emph {et~al.}(1987)\citenamefont
  {Leggett}, \citenamefont {Chakravarty}, \citenamefont {Dorsey}, \citenamefont
  {Fisher}, \citenamefont {Garg},\ and\ \citenamefont {Zwerger}}]{Leggett1987}%
  \BibitemOpen
  \bibfield  {author} {\bibinfo {author} {\bibfnamefont {A.~J.}\ \bibnamefont
  {Leggett}}, \bibinfo {author} {\bibfnamefont {S.}~\bibnamefont
  {Chakravarty}}, \bibinfo {author} {\bibfnamefont {A.~T.}\ \bibnamefont
  {Dorsey}}, \bibinfo {author} {\bibfnamefont {M.~P.~A.}\ \bibnamefont
  {Fisher}}, \bibinfo {author} {\bibfnamefont {A.}~\bibnamefont {Garg}}, \ and\
  \bibinfo {author} {\bibfnamefont {W.}~\bibnamefont {Zwerger}},\ }\href
  {\doibase 10.1103/RevModPhys.59.1} {\bibfield  {journal} {\bibinfo  {journal}
  {Rev. Mod. Phys.}\ }\textbf {\bibinfo {volume} {59}},\ \bibinfo {pages} {1}
  (\bibinfo {year} {1987})}\BibitemShut {NoStop}%
\bibitem [{\citenamefont {DiVincenzo}(1995)}]{DiVincenzo1995}%
  \BibitemOpen
  \bibfield  {author} {\bibinfo {author} {\bibfnamefont {D.~P.}\ \bibnamefont
  {DiVincenzo}},\ }\href {\doibase 10.1103/PhysRevA.51.1015} {\bibfield
  {journal} {\bibinfo  {journal} {Phys. Rev. A}\ }\textbf {\bibinfo {volume}
  {51}},\ \bibinfo {pages} {1015} (\bibinfo {year} {1995})}\BibitemShut
  {NoStop}%
\bibitem [{\citenamefont {Unruh}(1995)}]{Unruh1995}%
  \BibitemOpen
  \bibfield  {author} {\bibinfo {author} {\bibfnamefont {W.~G.}\ \bibnamefont
  {Unruh}},\ }\href {\doibase 10.1103/PhysRevA.51.992} {\bibfield  {journal}
  {\bibinfo  {journal} {Phys. Rev. A}\ }\textbf {\bibinfo {volume} {51}},\
  \bibinfo {pages} {992} (\bibinfo {year} {1995})}\BibitemShut {NoStop}%
\bibitem [{\citenamefont {Nolting}(2009)}]{Noltingmanybody}%
  \BibitemOpen
  \bibfield  {author} {\bibinfo {author} {\bibfnamefont {W.}~\bibnamefont
  {Nolting}},\ }\href {\doibase 10.1007/978-3-540-71931-1} {\emph {\bibinfo
  {title} {Fundamentals of Many-body Physics}}}\ (\bibinfo  {publisher}
  {Springer Berlin Heidelberg},\ \bibinfo {year} {2009})\BibitemShut {NoStop}%
\bibitem [{\citenamefont {Scheja}\ and\ \citenamefont
  {Storch}(1988)}]{SchejaStorchbook}%
  \BibitemOpen
  \bibfield  {author} {\bibinfo {author} {\bibfnamefont {G.}~\bibnamefont
  {Scheja}}\ and\ \bibinfo {author} {\bibfnamefont {U.}~\bibnamefont
  {Storch}},\ }\href@noop {} {\emph {\bibinfo {title} {Lehrbuch der Algebra}}}\
  (\bibinfo  {publisher} {Stuttgart : Teubner},\ \bibinfo {year}
  {1988})\BibitemShut {NoStop}%
\bibitem [{\citenamefont {Gray}(2006)}]{Graybook}%
  \BibitemOpen
  \bibfield  {author} {\bibinfo {author} {\bibfnamefont {R.~M.}\ \bibnamefont
  {Gray}},\ }\href@noop {} {\emph {\bibinfo {title} {Toeplitz and Circulant
  Matrices: A review}}},\ edited by\ \bibinfo {editor} {\bibfnamefont
  {S.}~\bibnamefont {Verdu}},\ \bibinfo {series} {Foundations and Trends in
  Communications and Information Theory}, Vol.~\bibinfo {volume} {2}\ (\bibinfo
   {publisher} {now Publishers Inc.},\ \bibinfo {year} {2006})\ pp.\ \bibinfo
  {pages} {155--239}\BibitemShut {NoStop}%
\bibitem [{\citenamefont {Whittaker}\ and\ \citenamefont
  {Watson}(1963)}]{Whittaker1963}%
  \BibitemOpen
  \bibfield  {author} {\bibinfo {author} {\bibfnamefont {E.~T.}\ \bibnamefont
  {Whittaker}}\ and\ \bibinfo {author} {\bibfnamefont {G.~N.}\ \bibnamefont
  {Watson}},\ }\href@noop {} {\emph {\bibinfo {title} {A Course of Modern
  Analysis}}},\ \bibinfo {edition} {4th}\ ed.\ (\bibinfo  {publisher}
  {Cambridge University Press},\ \bibinfo {year} {1963})\BibitemShut {NoStop}%
\bibitem [{\citenamefont {Pemberton}\ and\ \citenamefont
  {Rau}(2001)}]{Pemberton2001}%
  \BibitemOpen
  \bibfield  {author} {\bibinfo {author} {\bibfnamefont {M.}~\bibnamefont
  {Pemberton}}\ and\ \bibinfo {author} {\bibfnamefont {N.}~\bibnamefont
  {Rau}},\ }\href {http://books.google.com.au/books?id=rJ5d7DW2t4UC} {\emph
  {\bibinfo {title} {Mathematics For Economists}}}\ (\bibinfo  {publisher}
  {Manchester University Press},\ \bibinfo {year} {2001})\ \bibinfo {note}
  {chapter 13.4, "Quadratic Forms"p. 241 in second edition}\BibitemShut
  {NoStop}%
\bibitem [{\citenamefont {Jones}()}]{Jonesbook}%
  \BibitemOpen
  \bibfield  {author} {\bibinfo {author} {\bibfnamefont {F.}~\bibnamefont
  {Jones}},\ }\href@noop {} {\enquote {\bibinfo {title} {Honors calculus
  iii/iv},}\ }\bibinfo {howpublished} {Online book},\ \bibinfo {note} {chap. 4:
  Symmetric matrices and the second derivative test. pg. 25-29}\BibitemShut
  {NoStop}%
\bibitem [{\citenamefont {Bloch}(1957)}]{Bloch1957}%
  \BibitemOpen
  \bibfield  {author} {\bibinfo {author} {\bibfnamefont {F.}~\bibnamefont
  {Bloch}},\ }\href {\doibase 10.1103/PhysRev.105.1206} {\bibfield  {journal}
  {\bibinfo  {journal} {Physical Review}\ }\textbf {\bibinfo {volume} {105}},\
  \bibinfo {pages} {1206} (\bibinfo {year} {1957})}\BibitemShut {NoStop}%
\bibitem [{\citenamefont {Redfield}(1957)}]{Redfield1957}%
  \BibitemOpen
  \bibfield  {author} {\bibinfo {author} {\bibfnamefont {A.~G.}\ \bibnamefont
  {Redfield}},\ }\href {\doibase 10.1147/rd.11.0019} {\bibfield  {journal}
  {\bibinfo  {journal} {IBM Journal of Research and Development}\ }\textbf
  {\bibinfo {volume} {1}},\ \bibinfo {pages} {19 } (\bibinfo {year}
  {1957})}\BibitemShut {NoStop}%
\bibitem [{\citenamefont {Wangsness}\ and\ \citenamefont
  {Bloch}(1953)}]{Wangsness1953}%
  \BibitemOpen
  \bibfield  {author} {\bibinfo {author} {\bibfnamefont {R.~K.}\ \bibnamefont
  {Wangsness}}\ and\ \bibinfo {author} {\bibfnamefont {F.}~\bibnamefont
  {Bloch}},\ }\href {\doibase 10.1103/PhysRev.89.728} {\bibfield  {journal}
  {\bibinfo  {journal} {Physical Review}\ }\textbf {\bibinfo {volume} {89}},\
  \bibinfo {pages} {728} (\bibinfo {year} {1953})}\BibitemShut {NoStop}%
\bibitem [{\citenamefont {Rivas}\ and\ \citenamefont
  {Huelga}(2012)}]{Rivasbook}%
  \BibitemOpen
  \bibfield  {author} {\bibinfo {author} {\bibfnamefont {A.}~\bibnamefont
  {Rivas}}\ and\ \bibinfo {author} {\bibfnamefont {S.~F.}\ \bibnamefont
  {Huelga}},\ }\href {http://books.google.com.au/books?id=FGCuYsIZAA0C} {\emph
  {\bibinfo {title} {Open Quantum Systems: An Introduction}}},\ Springerbriefs
  in Physics\ (\bibinfo  {publisher} {Springer},\ \bibinfo {year}
  {2012})\BibitemShut {NoStop}%
\end{thebibliography}%
\end{document}